\documentclass[10pt,conference]{IEEEtran}
\usepackage{graphicx}
\usepackage{hyperref}
\usepackage{listings}
\usepackage{cite}
\usepackage{tikz}
\usepackage{multirow}
\usepackage{algpseudocode}
\usepackage{algorithm}
\usepackage{float}
\usepackage{mdframed}
\usetikzlibrary{shapes.geometric, arrows.meta, positioning}
\nocite{*}

\tikzstyle{decision} = [diamond, draw, fill=blue!10, text width=5cm, align=center, inner sep=1pt]
\tikzstyle{terminal} = [circle, draw, fill=blue!10, minimum size=2cm, align=center]
\tikzstyle{arrow} = [thick, ->, >=stealth]

\newcommand{\lastmerge}{\textsc{LastMerge}}

\newcommand{\mergiraf}{\textit{Mergiraf}}
\newcommand{\jdime}{\textit{jDime}}
\newcommand{\spork}{\textit{Spork}}
\newcommand{\treesitter}{\textit{Tree Sitter}}
\newcommand{\afp}{\textit{aFP}}
\newcommand{\afn}{\textit{aFN}}
\newcommand{\afps}{\textit{aFPs}}
\newcommand{\afns}{\textit{aFNs}}

\title{\lastmerge{} – A language-agnostic structured tool for code integration}
\author{João Pedro Duarte, Paulo Borba, Guilherme Cavalcanti}
\begin{document}

\maketitle

\begin{abstract}
Unstructured line-based merge tools are widely used in practice. 
Structured AST-based merge tools show significantly improved merge accuracy, but are rarely used in practice because they are language specific and costly, consequently not being available for many programming languages.
To improve merge accuracy for a wide range of languages, we propose \lastmerge{}, a \emph{generic} structured merge tool that can be configured through a thin interface that significantly reduces the effort of supporting structured merge.
To understand the impact that \emph{generic} structured merge might have on merge accuracy and performance, we run an experiment with four structured merge tools: two Java specific tools, \jdime{} and \spork{}, and their \emph{generic} counterparts, respectively \lastmerge{} and \mergiraf{}.
Using each tool, we replay merge scenarios from a significant dataset, and collect data on runtime, behavioral divergences, and merge accuracy. 
Our results show no evidence that \emph{generic} structured merge significantly impacts merge accuracy. 
Although we observe a difference rate of approximately 10\% between the Java specific tools and their \emph{generic} counterparts, most of the differences stem from implementation details and could be avoided. 
We find that \lastmerge{} reports 15\% fewer false positives than \jdime{} while \mergiraf{} misses 42\% fewer false negatives than \spork{}.
Both \emph{generic} tools exhibit comparable runtime performance  to the state of the art language specific implementations.
These results suggest that \emph{generic} structured merge tools can effectively replace language-specific ones, paving the way for broader adoption of structured merge in industry.
\end{abstract}
\section{Introduction}

In most projects, developers collaborate by working on separate branches or repositories (like local and remote ones), and later merge their changes into the main codebase.
Version Control Systems (VCSs) typically rely on line-based, \emph{unstructured}, merge tools such as \textit{diff3}~\cite{mens2002}. 
These tools compare file revisions based solely on the textual content of their lines, without considering the syntactic or semantic structure of the code~\cite{sanjeev2007}. 
Such unstructured merge techniques are fast, language-agnostic, and widely used in practice.
However, they often produce spurious conflicts that waste developer effort to fix semantically harmless issues that could be otherwise automatically resolved. 
At the same time, unstructured merge may overlook actual conflicts, which can silently propagate into final artifacts and cause regressions in production.

To improve the accuracy of merging software revisions, researchers have proposed tools that leverage the syntactic structure of the source code~\cite{hunt2002,apel2011,fengmin2019,westfechtel1991,clementino2021,buffenbarger1995,larsen2023,apel2012}. 
Unlike unstructured tools, \emph{structured} merge tools parse the source code into an Abstract Syntax Tree (AST) and apply tree matching and combination algorithms to generate the merged artifact. 
Prior studies have shown that structured merge significantly improves merge accuracy compared to unstructured techniques~\cite{seibt2021,schesch2024}. 
Despite these advances, structured tools remain largely absent from current practice especially for two reasons. 
First, as they strongly rely on the syntax and semantics of specific programming languages, structured tools proposed so far are language specific; 
substantial implementation and maintenance effort is needed for supporting a new   language. 
Second, for being costly, not many languages are supported by structured tools, which  hinders adoption by developers who routinely work with multiple languages, as often needed in practice.

To reduce these barriers and improve merge accuracy for a wide range of programming languages, we propose \lastmerge{}, a \emph{generic} structured merge tool that can be easily configurable for each language.
It relies on a core merge engine that operates over generic trees produced by an extensible and fast parser framework~\cite{treesitter} that has been instantiated for more than 350 languages, and is in production at GitHub.
By feeding the engine with a high level description of language specific aspects (such as node labelling and restrictions to permutation of node children) that are known to be relevant for structured merge, developers can easily adapt \lastmerge{} for new languages, or refine support for existing ones.
This thin configuration interface significantly reduces the effort of having structured merge for a wide range of languages.

To understand the impact that a \emph{generic} structured merge technique might have on merge accuracy and computational performance, we run an experiment with four structured merge tools: 
\jdime{}~\cite{apel2012}, a well-known Java specific tool, and \lastmerge{}, which adopts a similar algorithm but on top of a language independent AST and \emph{generic} setting;
and \spork{}~\cite{larsen2023}, a more recent Java specific tool, and \mergiraf{},\footnote{\url{https://mergiraf.org/}} which adopts a similar algorithm but on top of the same language independent setting as \lastmerge{}. 
We replay merge scenarios using each tool (the generic ones instantiated with Java syntactic and semantic details) in a significant dataset~\cite{schesch2024}, and collect data on runtime, behavioral divergences, and merge accuracy. 
Specifically, we compute the number of spurious conflicts (false positives) and actual missed conflicts (false negatives). 
We address the following research questions: How \emph{generic} structured merge impacts merge accuracy? How \emph{generic} structured merge impacts merge runtime performance?

Our results show no evidence that \emph{generic} structured merge significantly impacts merge accuracy. 
Although we observe a difference rate of approximately 10\% between the Java specific tools (\jdime{} and \spork{}) and their \emph{generic} counterparts (\lastmerge{} and \mergiraf{}) instantiated for Java, most of the differences stem from implementation details and configuration choices, not from design decisions implied by the generality requirement. 
We also find that \lastmerge{} reports 15\% fewer false positives than \jdime{}, while \mergiraf{} misses 42\% fewer false negatives than \spork{}.
Furthermore, both \emph{generic} tools exhibit comparable runtime performance  to the state of the art language-specific implementations. 
These results suggest that \emph{generic} structured merge tools can effectively replace language-specific ones, achieving similar levels of accuracy and efficiency, thus paving the way for broader adoption of structured merge in industry.

\section{Unstructured and Structured Merge}
\label{sec:motivating}

\begin{figure}
    \centering
    \includegraphics[width=0.49\textwidth]{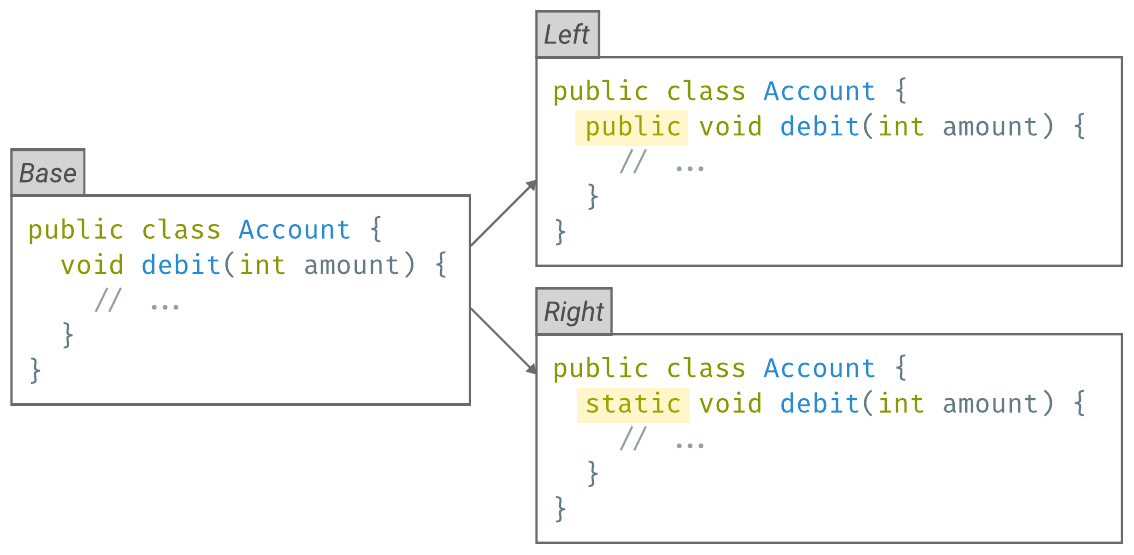}
    \caption{An example of a merge scenario. \textit{Base} shows the initial revision, or the shared common ancestor. Both \textit{Left} and \textit{Right} are revisions that independently introduce changes to \textit{Base}. Changes are highlighted in yellow.}
    \label{fig:motivating}
\end{figure}

To illustrate the differences between unstructured and structured merge, and the costs of supporting the latter for a number of languages, consider the merge scenario illustrated in Figure~\ref{fig:motivating}. 
It shows the initial declaration of the \texttt{debit} method in the \texttt{Account} class. Starting from this base version, two developers, \textit{Left} and \textit{Right}, independently modify the method. 
\textit{Left} makes the method \texttt{public}, while \textit{Right} makes the method \texttt{static}. 
As these changes differ, they must be merged to produce the final version of the \texttt{Account} class.

\begin{figure}
    \centering
    \includegraphics[width=0.4\textwidth]{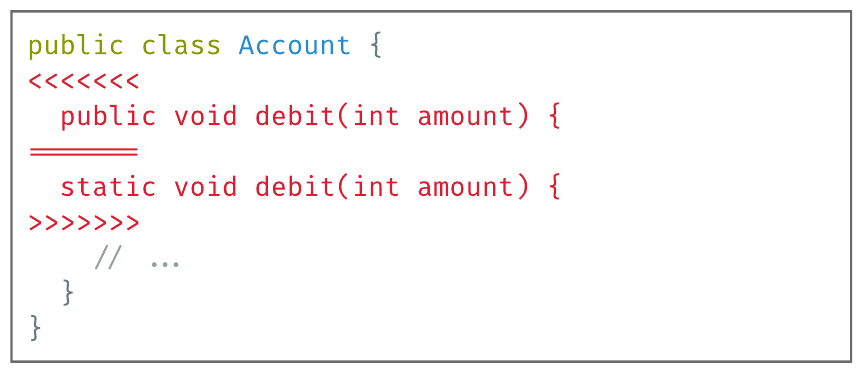}
    \caption{The output of unstructured merge. Since it only relies on the textual content of the lines, a conflict between the changes introduced by \textit{Left} and \textit{Right} is reported, as presented in red.}
    \label{fig:motivating_unstructured}
\end{figure}

In this scenario, an unstructured merge tool such as \textit{diff3} performs a line-based comparison between the revisions, using the common base version as a reference. 
Since both developers modify the same line, the tool is unable to integrate the changes and reports a conflict, as illustrated in Figure~\ref{fig:motivating_unstructured}. 
Resolving this conflict requires manual intervention to combine the modifications to preserve developers intentions.

\begin{figure}[H]
    \centering
    \includegraphics[width=0.4\textwidth]{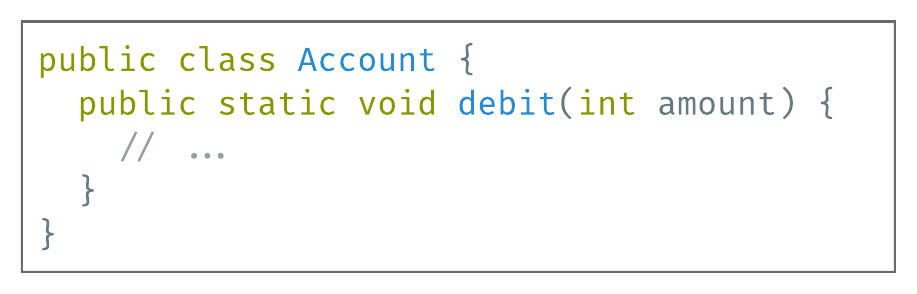}
    \caption{The output of structured merge. Since the changes made by \textit{Left} and \textit{Right} occur in different tree nodes, no conflict is reported.}
    \label{fig:motivating_structured}
\end{figure}

As shown in Figure~\ref{fig:motivating_structured}, structured merge correctly merges \textit{Left} and \textit{Right} changes avoiding developer effort.
Structured tools~\cite{hunt2002,apel2011,fengmin2019,westfechtel1991,clementino2021,buffenbarger1995,larsen2023,apel2012} are language-specific and leverage language syntax and semantics.
In the Java case, they explore the fact that modifiers such as \texttt{public} and \texttt{static} can be applied together and in any order. 
Instead of comparing lines of text, these tools firstly construct tree representations of the source code--- typically Abstract Syntax Trees (ASTs)--- for each revision to be merged. 
During the \textit{matching} phase, the tool correlates common and modified nodes across revisions. 
In the subsequent \textit{amalgamation} phase, it merges the nodes based on the collected matching information. 
Conflicts are reported only when different changes affect corresponding tree nodes. 
Structured merge also avoids spurious conflicts when, for instance, developers independently add declarations (field, method, etc.) to the same area of the text, or change syntactically separate parts of an expression or statement, even if they appear in the same or consecutive lines of code.

Although structured merge outperforms unstructured merge by reporting fewer spurious conflicts~\cite{seibt2021,schesch2024}, and even detecting conflicts that are missed by unstructured tools, creating such a tool demands significant effort for each language that needs to be supported. 
Besides creating or adapting\footnote{Tools often reuse existing parser infrastructure: \spork{} relies on Spoon \cite{larsen2023}, and \jdime{} on JastAddJ \cite{apel2012}.} language-specific parsers and ASTs, one has to implement the whole merging (matching, amalgamation, etc.) engines for each AST, which is often expensive. 
Maintaining such engines is also costly, as they need to be fixed or updated for each supported language.  
This explains why structured merge tools are available for only a few languages, and have not been widely adopted in industry, where tools that support multiple languages are often needed for most nontrivial projects.
\section{Generic Structured Merge}
\label{sec:language_independent_structured_merge}
To reduce the problems discussed in the previous section, and improve merge accuracy for a wide range of programming languages, we propose \lastmerge{} (\textit{Language Agnostic Structured Tool for Code Merging}), a \emph{generic} structured merge tool that can be easily configurable for each language.
Here we explain \lastmerge{}'s main design decisions, as well as a high level view of the ones chosen for \mergiraf{}, the other \emph{generic} structured tool we use in our experiment to understand whether our evaluation results are specific to \lastmerge{} or generalize beyond our particular design choices.

\subsection{\lastmerge{}}
\label{sec:last_merge}

\begin{figure}
    \centering
    \includegraphics[width=0.4\textwidth]{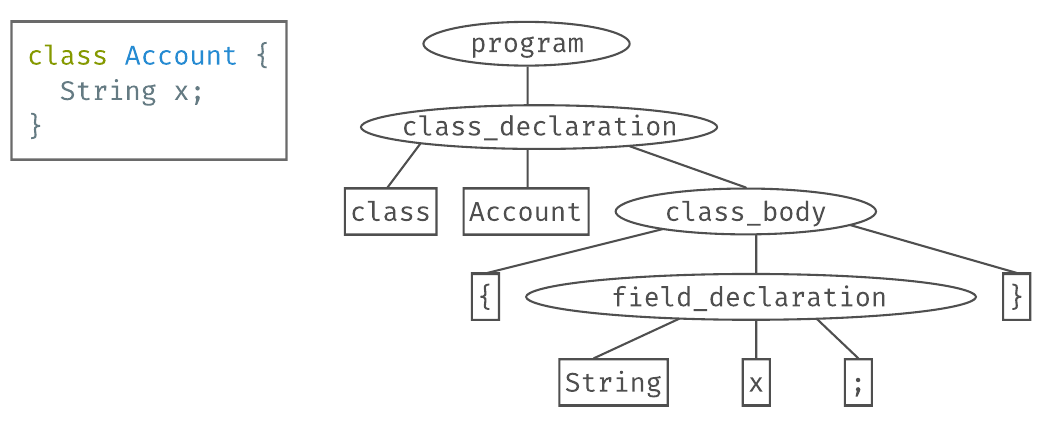}
    \caption{An CST produced by \treesitter. An CST represents all the syntactic information of the original source code, including lexical elements.}
    \label{fig:motivating_cst_vs_ast}
\end{figure}

To achieve generality, \lastmerge{} relies on a core merge engine that operates over generic trees produced by \treesitter{}~\cite{treesitter}, an extensible and fast parser framework that has been instantiated for more than 350 languages, and is in production at GitHub.
\treesitter{} allows users to define a Context-Free Grammar (CFG) using a domain-specific language (DSL) to generate a parser. 
This parser builds a Concrete Syntax Tree (CST), a tree representation of the source code that preserves all syntactic elements; nodes are represented as either \textit{Terminal} (leaf nodes, such as literals) or \textit{NonTerminal} (internal nodes, such as method declarations).
Unlike Abstract Syntax Trees (ASTs), CSTs retain more granular and less abstract information, as illustrated in Figure~\ref{fig:motivating_cst_vs_ast}.
Although several parser generators exist, the main advantage of \treesitter{} lies in its extensive collection of community maintained grammars for most programming languages used in industry. 
This ecosystem enables developers to build tools that aim to be language independent by focusing on a core that operates over generic tree nodes, while delegating the parsing to \treesitter{}.

\begin{figure}
    \centering
    \includegraphics[width=0.5\textwidth]{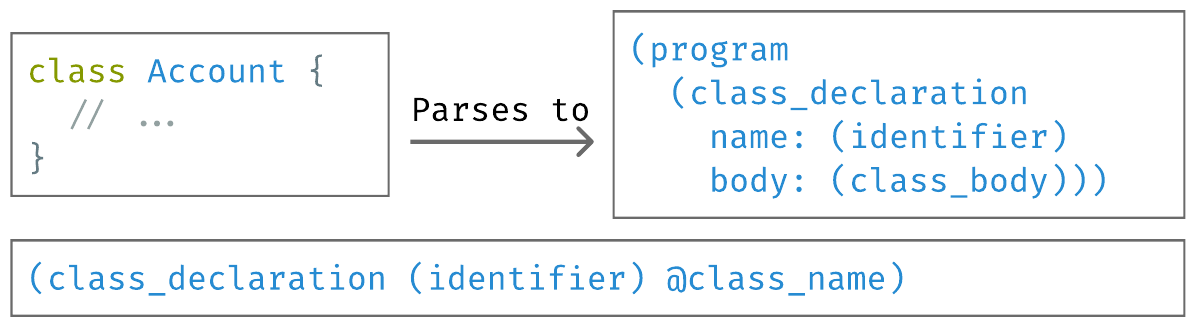}
    \caption{On top a class declaration in Java and its \treesitter{} parsed version. On the bottom, a \treesitter{} query. Queries consist of one or more \textit{patterns} that are specified as \textit{S-Expressions}. In this example, the named capture \textit{class\_name} holds the identifier (name) of the class \texttt{Account}.}
    \label{fig:last_merge_query}
\end{figure}

By plugging a \treesitter{} parser, and feeding \lastmerge{}'s merge engine with a high level description of language specific aspects (such as which \textit{NonTerminal} nodes can be treated as unordered) that are known to be relevant for structured merge, developers can easily adapt \lastmerge{} for new languages, or refine support for existing ones.
This thin configuration interface significantly reduces the effort of having structured merge for a wide range of languages.
The user also has to instruct the tool on how to extract node identifiers. 
This is specified using \treesitter{} queries, which are executed during parsing to locate specific code structures within nodes through pattern matching. 
Figure~\ref{fig:last_merge_query} illustrates an example \treesitter{} representation of a class declaration alongside a query that extracts the class name, which can be used to uniquely identify it within the context of a Java file.

\begin{figure}[H]
    \centering
    \includegraphics[width=0.5\textwidth]{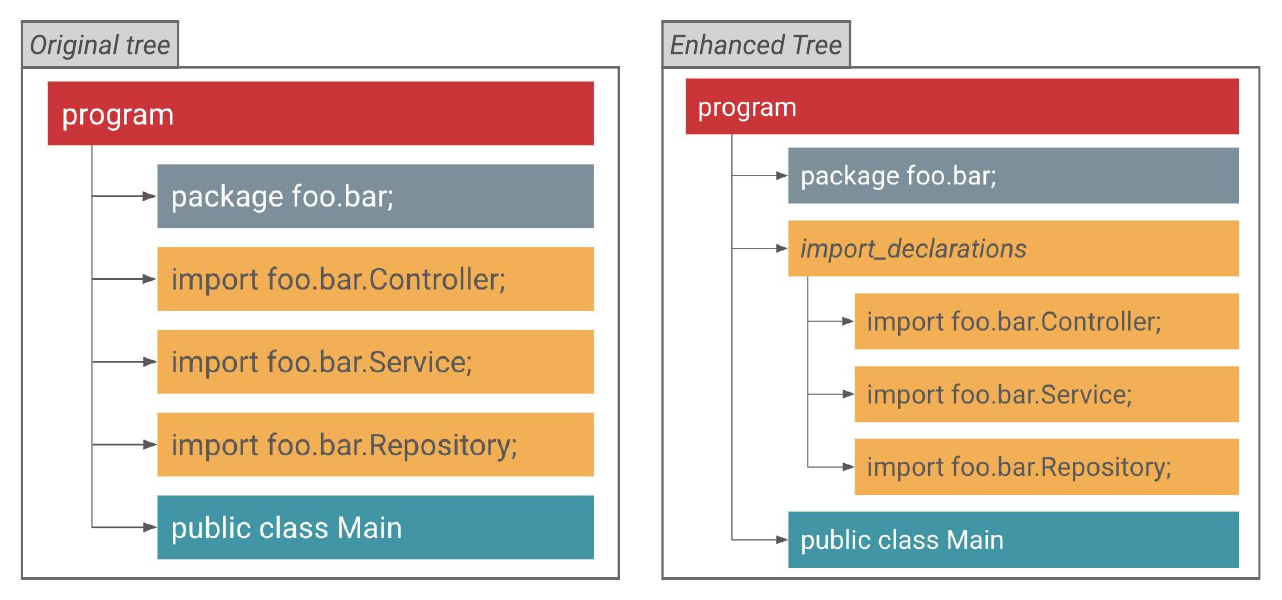}
    \caption{A transformation applied over a tree by a parsing handler. On the left, the original tree; on the right, the final version. This parsing handlers operates by grouping import declarations into a single \textit{NonTerminal} node.}
    \label{fig:parsing_handlers_last_merge}
\end{figure}

It is also possible to hook in after the parsing stage by implementing \textit{parsing handlers}, which execute language-specific procedures to augment the original parse tree and improve merge accuracy. 
For example, we implemented a parsing handler for Java that groups import declarations into a single \textit{NonTerminal} unordered node, as shown in Figure~\ref{fig:parsing_handlers_last_merge}. 
This reorganization allows the tool to apply unordered matching specifically to import declarations, making it possible to match such declarations across common use cases such as import reordering, thus resulting in higher matching accuracy and improved merge outcomes.

Architecturally, \lastmerge{} follows a sequential pipeline composed of three main steps: \textit{parsing}, \textit{matching} and \textit{merging}.

\subsubsection{Parsing}
\label{sec:last_merge_parsing}

\lastmerge{} relies on \treesitter{} to convert each revision into a Concrete Syntax Tree (CST).
\textit{NonTerminal} nodes can be marked as \textit{Unordered}, indicating that their children can be safely permuted without affecting program semantics. 
Nodes may also include an optional identifier used to uniquely distinguish them among their siblings. Both identifier assignment and children ordering are language-specific aspects that must be properly configured to ensure correct behavior.

To better illustrate these concepts, consider again the example in Figure \ref{fig:motivating}. In this scenario, the order of method declaration modifiers, such as \texttt{public} and \texttt{static}, can be altered freely, as they are marked as unordered nodes, without affecting the program semantics.
Additionally, the signature of the method \texttt{debit(int amount)} can be used to uniquely identify it among the children of the \texttt{Account} class.

\subsubsection{Matching}
\label{sec:last_merge_matching}

This and the next step are based on \jdime{}'s algorithms~\cite{apel2012}, which were adapted to work in a generic language-independent tree. 
Matchings between each pair of revisions--- (\emph{base}, \emph{left}), (\emph{base}, \emph{right}), and (\emph{left}, \emph{right})--- are first computed by associating nodes in each  revision in a pair with corresponding nodes in the other revision. 
The matching process disallows pairing a \textit{Terminal} node with a \textit{NonTerminal} node.
Moreover, even when nodes share the same type (\textit{NonTerminal} or \textit{Terminal}), they can only be matched if their kinds (method declaration, literal, etc.) are identical.
This restriction ensures the merge algorithm only matches semantically equivalent nodes. 

The matching algorithms depend heavily on the types of nodes being compared. 
Matching \textit{Terminal} nodes, for example, is relatively straightforward; a match is assigned only if their values are identical.
Matching \textit{NonTerminal} nodes is more complex.
When their root nodes match--- either because they share the same identifier or kind--- we recursively compute matches between each pair of children to find a maximum matching. 
This process occurs level-wise, so nodes only match with others siblings at the same tree level.
The specific algorithm to compute this maximum matching varies depending on whether these \textit{NonTerminal} nodes are ordered or unordered.

For nodes with ordered children, we use Yang's algorithm~\cite{yang1991} to compute the maximum matching in quadratic time. 
The algorithm uses a dynamic programming approach to find the number of pairs in a maximum matching between the trees. 
These maximum matching candidates are computed by recursively traversing both trees and computing the maximum matching between each pair of their children~\cite{yang1991}.
If nodes have unordered children, we rely on a linear programming approach that runs in cubic time to find the maximum matching, similar to \jdime{}'s implementation~\cite{apel2012}. 

\subsubsection{Merging}
\label{sec:last_merge_merging}

To construct the final merged version, the algorithm traverses pairs of nodes from each parent in a depth-first manner. 
It leverages the previously computed matching information to decide which nodes to retain or discard, how to integrate concurrent changes to the same node, and when to report conflicts--- for example, when one revision modifies a node that the other has removed.
Depending on the node type, certain heuristics can be applied to resolve conflicts that unstructured techniques cannot handle; such as reordering children of an unordered \textit{NonTerminal} node.

\subsection{\mergiraf{}}

Similarly to \lastmerge{}, \mergiraf{} is based on \treesitter{}'s parse infrastructure and language-independent tree. 
However, instead of adapting \jdime{}'s algorithms to a generic context, \mergiraf{} opts for adapting \spork{}'s algorithms.
By design, \mergiraf{} explores auto tuning~\cite{apel2012}, first attempting unstructured merge of its input files. 
Only if conflicts arise during this attempt, it resort to structured merge.

When using structured merge, the tool builds fictional trees upon the conflicts found during the line-based merge of the file. 
This allows \mergiraf{} to aggressively pre-assign matchings between the revisions, which significantly speeds up the matching process. 
The remaining matchings between each pair of revisions are computed using only the GumTree algorithm~\cite{gumtree}, differently from \lastmerge{} that combines different algorithms.
Finally, revisions are merged into the final artifact using the same algorithm used by \spork{}~\cite{larsen2023}.

\section{Evaluating Generic Structured Merge}
\label{sec:empirical_study}

To understand the impact that a \emph{generic} structured merge technique might have on merge accuracy and computational performance, we run an experiment with generic tools (which rely on  matching and merging language independent trees) and their language specific counterparts. 
We pair \lastmerge{} with \jdime{}, and \mergiraf{} with \spork{}, reflecting the influence of existing state of the art, language specific, tools on the design and implementation of the generic ones. 

This pairing helps to isolate the generality aspect, mitigating bias that could arise from the effect of design decisions (algorithms, etc.) not related to implementing the generic requirement. 
This setting, especially with two generic tools, also helps us to investigate whether results are consistent across two state of the art algorithms for structured merge.

We replay merge scenarios (quadruples formed by a merge commit, its two parents, and a base commit) using each tool (the generic ones instantiated with Java syntactic and semantic details), and collect data on runtime, behavioral divergences, and merge accuracy. 

Here, we present the research questions we address, our sampling process, and further details of our experiment design. 

\subsection{Research Questions}
\label{sec:research_questions}

With the goal of understanding whether \emph{generic} structured merge tools can effectively replace language-specific ones, achieving similar levels of accuracy and efficiency, we ask the following research questions:
\subsubsection*{\textbf{RQ1:} How \emph{generic} structured merge impacts merge accuracy?}

To address this question, we compute the number of spurious conflicts (false positives) and actual missed conflicts (false negatives), but we do that comparatively, only when the pair of tools being compared yield different results.
So if both tools erroneously report a conflict, we do not consider that as a false positive in our analysis.
We pay only attention to scenarios where one tool reports a conflict and the other does not, for instance. 
This is needed because establishing sound conflict ground truth for a large sample is hard.
So we adopt the concepts of \textit{added false positives (\afps{})} and \textit{added false negatives (\afns{})}~\cite{cavalcanti2017,cavalcanti2024}, which are explained in detail latter.
To answer this question, we also manually analyze a number of cases to understand whether differences in conflict detection accuracy occur due to programming language independence or other factors.

\subsubsection*{\textbf{RQ2:} How \emph{generic} structured merge impacts merge runtime performance?}

To answer this research question, we measure the runtime of each tool in every merge scenario. Pairwise comparisons enable a clearer understanding of how different tree structures affect the performance of structured merge tools. 
We also examine whether programming language independence imposes a prohibitively high performance cost.

By answering these questions we hope to understand whether generic structured tools such as \lastmerge{} have the potential to pave the way for broader adoption of structured merge in industry, as they can be easily adapted to multiple languages.

\subsection{Sampling}

For comparing the merge tools, we use the dataset of merge scenarios published by Schesch et al.~\cite{schesch2024}. 
The sample consists of 5,983 merge scenarios from 1,116 open source projects. 
Projects are extracted from GitHub's Greatest Hits~\cite{GitHubGreatestHits} and Reaper~\cite{munaiah2017} datasets, and were carefully filtered so that users can rely on significant buildable Java projects that are relevant within the open-source community.
The dataset includes scenarios with non-trivial test suites that pass on both parents within a specified timeout.
This is particularly useful because it allows us to rely on test execution to check merge accuracy; if tools yield different results but project tests pass in the results of just one of the tools, we know the other tool has a problem.

When trying to replicate the original dataset, a number of scenarios could not be retrieved due to external factors--- such as when the GitHub repository is unavailable.
We discard these in our study.
We further filter the sample to ensure each scenario contains at least one file that was mutually modified by both parents.
This excludes scenarios where merging is trivially achieved by selecting the revision that introduces the changes; the tools would yield the same results for these scenarios, not contributing to our comparative analysis. 
Additionally, we remove scenarios in which any of the evaluated merge tools crashed during execution.
As a consequence of these filtering, we perform our experiment on a sample of 5,229 scenarios, spanning 13,675 mutually modified files.

\subsection{Checking Merge Accuracy and Performance}
\label{sec:methodology}

To answer our research questions, we replay merge scenarios with the four tools, and collect metrics on runtime performance and differences in merge results among them.

To answer RQ1, following the merge process, we collect metrics to estimate merge accuracy, as explained in Section~\ref{sec:research_questions}.  
In particular, we adopt a relative comparison~\cite{cavalcanti2017,cavalcanti2024}, computing the occurrence of false positives and false negatives of one tool \textit{in addition} to the other tool in the same pair. 
Given a pair of tools $(A, B)$, we say tool $A$ suffers an \textit{added false positive (\afp{})} in a scenario if it reports an spurious conflict that tool $B$ does not.
Similarly, $A$ suffers an \textit{added false negative (\afn{})} in a scenario if it fails to report an actual conflict detected by $B$.

\begin{figure*}
    \centering
    \includegraphics[width=.75\textwidth]{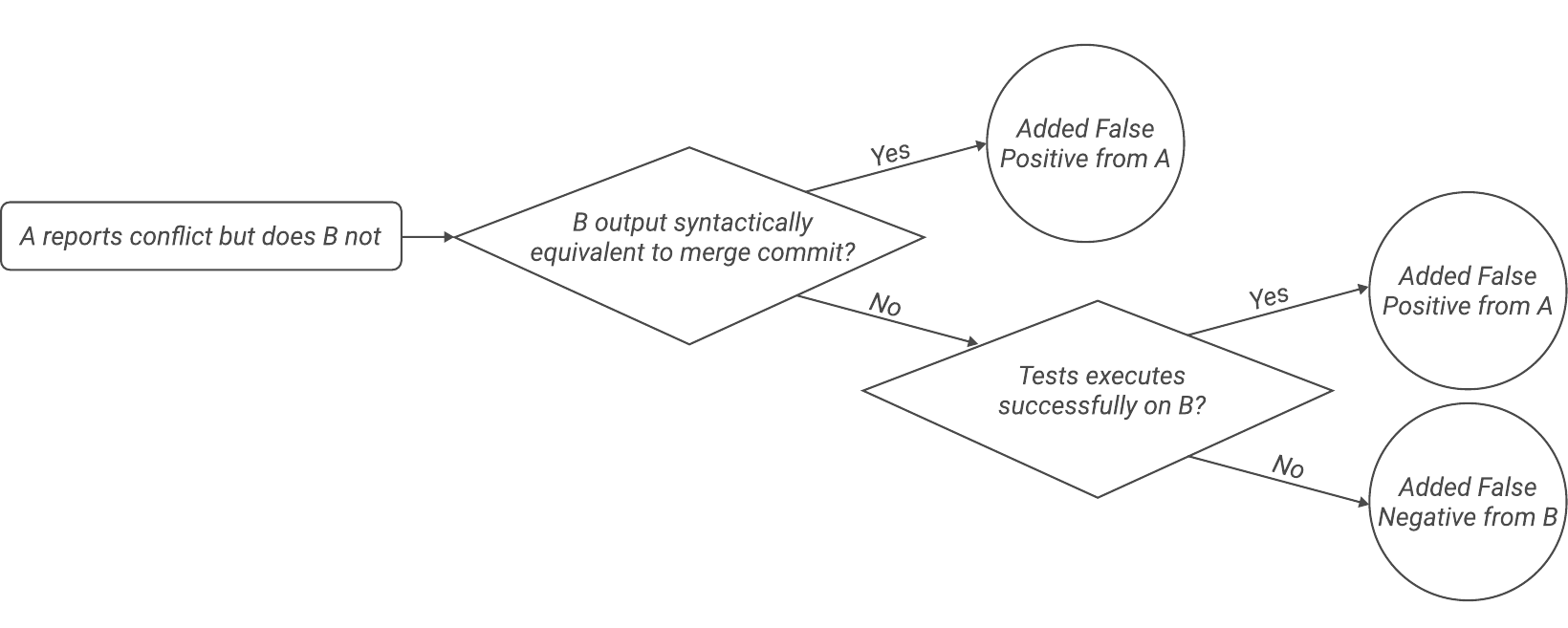}
    \caption{Deciding whether a tool has an added false positive (\afp{}) or an added false negative (\afn{}) in a merge scenario (merge commit in the repository, its parents, and their common ancestor). The output of $B$ is first compared to the merge commit in the repository. In summary, in the scenario that $A$ reports a conflict that $B$ does not, we check whether $B$ successfully and accurately resolves these conflicts.}
    \label{fig:afps_afns}
\end{figure*}

Note that our analysis metrics assumes both tools disagree on the presence of conflicts in a scenario.
This aligns with our goal to compare tools relatively, focusing not on the absolute number of conflicts each tool reports but on how they differ. 
Remember we are particularly interested in understanding whether the \emph{generic} nature of a tool impacts its accuracy in comparison to its language-specific counterpart.
Moreover, scenarios with disagreement are critical for developers, as incorrect conflict detection can result in faulty merges that are costly to fix. 
We calculate the number of \afps{} and \afns{} for a pair of tools $(A, B)$ by combining syntactic and semantic approaches, as illustrated in Figure~\ref{fig:afps_afns}.

Given a scenario where tool $A$ reports a conflict and $B$ does not, we aim to determine whether $B$ has successfully and accurately resolved the conflict. 
We begin by checking if the output produced by $B$ is syntactically equivalent to the merge commit of the original project repository, as developers are presumably happy with such result. 
To do so, we use \lastmerge{} to parse the files into trees and later rely on its matching algorithm to verify whether their program root nodes match entirely.
But before that, we need a normalization step.
Due to the use of more abstract AST representations, \jdime{} and \spork{} are known to modify the original source code during pretty-printing by inserting or removing tokens. 
These modifications can interfere with the syntactic equivalence check, as they alter the syntactical structure of the output files. 
To mitigate this issue, we first normalize formatting by running \jdime{} and \spork{} on the outputs produced by \lastmerge{} and \mergiraf{}, respectively. 
We then use the resulting normalized files for syntactic comparison. 
If the files are found to be syntactically equivalent, we assume $B$ correctly resolved the conflict reported by $A$, and classify the scenario as an \afp{} for $A$.

Conversely, if $B$'s output is not syntactically equivalent, we rely on executing the project test suite on the output generated by $B$. 
If all tests pass, we assume that the changes introduced by both parents are non-conflicting and that $B$ successfully integrated them. 
Since $A$ reported a conflict and thus failed to integrate the changes, we classify this as an \afp{} for $A$. 
However, if the tests fail, we assume $B$ did not correctly detect the conflict that $A$ detected, thus resulting in an \afn{} for $B$.

Finally, as a last step to answer RQ1, we conduct a manual code analysis to better understand the reasons behind tool differences. 
We randomly select 5 scenarios with \afps{} and 5 with \afns{} for each tool, resulting in a total of 40 scenarios. 
For each scenario, we investigate whether the observed differences in the tools outputs are due to programming language independence or from other factors, such as configuration differences or implementation details. 

To answer RQ2, we collect runtime execution metrics by measuring the time each tool takes to merge each scenario. 
To do so, we aggregate the execution times across all files within a scenario. 
To minimize the influence of external factors, each tool is executed sequentially ten times per file, with runtime measured in each run. 
We discard the first measurement--- used as a warm-up--- and compute the average of the remaining nine runs to determine the runtime for each file.

We provide the scripts and data associated with this study in our online appendix.\footnote{\url{https://anonymous.4open.science/r/experiment-last-merge-3908}}

\section{Results and discussion}
\label{sec:results_and_discussion}

In this section we present and discuss our findings, structured according to the research questions outlined in Section~\ref{sec:research_questions}, and the two pairs of $(\mathit{specific}, \mathit{generic})$ structured merge tools compared: $(\jdime{}, \lastmerge{})$ and $(\spork{}, \mergiraf{})$.

\subsection{How \emph{generic} structured merge impacts merge accuracy? (RQ1)}

As explained in Section~\ref{sec:methodology}, our analysis is comparative.
For each tool pair, our analysis focuses on scenarios in which the tools disagree on the existence of conflicts, as we are not concerned with cases where both tools fail or where both successfully perform the merge.
So we first summarize in Table~\ref{tab:merge_agreement_summary} the agreement and disagreement on the existence of conflicts between each pair of merge tools. 
We observe a disagreement rate on the existence of conflicts of 7.53\% between \lastmerge{} and \jdime{}, and of 12.22\% between \mergiraf{} and \spork{}. 
So, based on our sample, we observe the \emph{generic} structured merge tools behaving differently than their language-specific counterparts in a minor, but considerable, part of cases.

\begin{table}[H]
    \caption{Agreement rate on conflict existence in the analyzed merge scenarios. The total sum in each column can vary because not all scenarios were successfully integrated by each tool.}
    \centering
    \begin{tabular}{ccc}
        \hline
        \textbf{Situation} & \textbf{\jdime{} vs \lastmerge{}} & \textbf{\spork{} vs \mergiraf} \\
        \hline
        Agreement on \\ existence of & 4909 (92.5\%) & 4697 (88.7\%) \\
        conflicts & & \\
        \hline
        Disagreement on \\ existence of & 400 (7.5\%) & 601 (11.3\%) \\
        conflicts & & \\
        \hline
    \end{tabular}
    \label{tab:merge_agreement_summary}
\end{table}

In the following, we discuss how these behavior differences impact the accuracy of the tools to detect and resolve conflicts. 
Furthermore, we examine whether these differences are related to the generic aspects (language independent trees and algorithms) of both \lastmerge{} and \mergiraf{}.
We do this first for the pair \lastmerge{} and \jdime{}, and then for the pair \mergiraf{} and \spork{}.



\subsection*{\lastmerge{} and \jdime{}}

Table~\ref{tab:last_merge_jdime_afps_afns} shows the results of the analysis of the accuracy of \lastmerge{} and \jdime{} in terms of \afps{} and \afns{}. 
The results indicate that \lastmerge{} reports fewer \afps{}, but exhibits nearly three times more \afns{} than \jdime{}. 
In absolute terms, this corresponds to 56 extra \afns{}, which is considerable.
Proportionally to the number of scenarios analyzed, or even to the number of scenarios in which the tools differ, we observe less significant numbers (1.1\% and 14\%, respectively), which are nevertheless further considered for our manual analysis.

\begin{table}[htpb]
    \caption{Comparison of added false positives (\afps{}) and added false negatives (\afns{}) between \lastmerge{} and \jdime{}.}
    \centering
    \begin{tabular}{ccc}
        \hline
        \textbf{Situation} & \textbf{\jdime{}} & \textbf{\lastmerge} \\
        \hline
        Added false positives (\afps{}) & 153 & 130 \\
        \hline
        Added false negatives (\afns{}) & 29 & 85 \\
        \hline
    \end{tabular}
    \label{tab:last_merge_jdime_afps_afns}
\end{table}

As explained in more detail in the rest of the section, our manual analysis suggests that the observed discrepancies are primarily due to implementation details and configuration differences, rather than to \lastmerge{} relying on language independent trees and algorithms. 
Although \lastmerge{} algorithms borrow from \jdime{} exactly to reduce confounding factors, they rely on substantially different trees, and minor differences in algorithm implementations are expected when involving different developers and programming languages, as in this case.
More important, configuration details, especially in the Java configuration of structured tools, can lead to differences that can easily eliminated by adjusting the configuration.  

\begin{figure}
    \centering
    \includegraphics[width=0.5\textwidth]{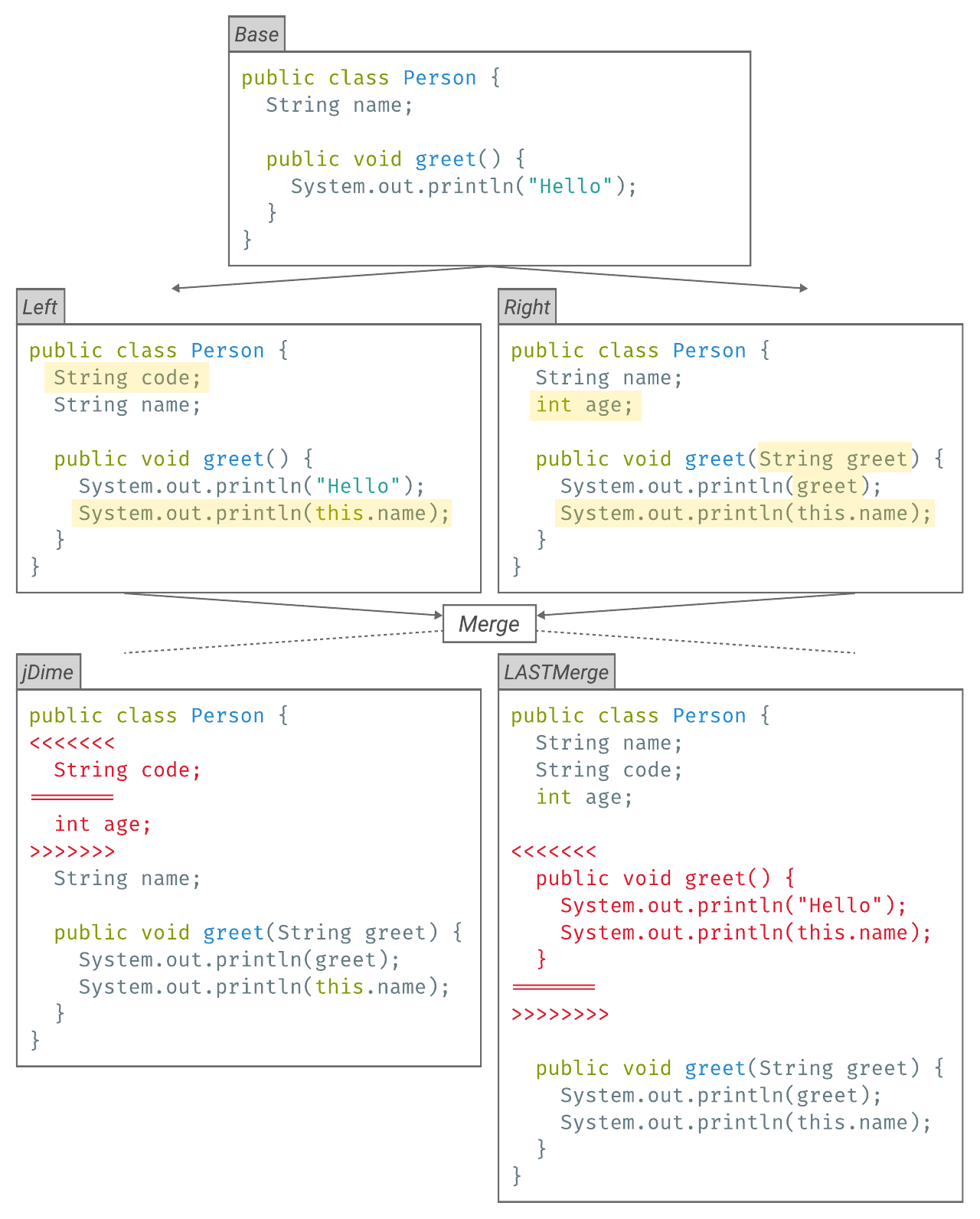}
    \caption{Merge scenario illustrating the differences in conflict detection between \lastmerge{} and \jdime{}. Changes are highlighted in yellow.}
    \label{fig:afp_jdime_last_merge}
\end{figure}

Starting with false positives, we found that in \jdime{} three out of five analyzed scenarios occur due to inadequate tool configuration. 
To illustrate this, consider the example in Figure~\ref{fig:afp_jdime_last_merge}. 
Starting from the \textit{Base} revision, \textit{Left} adds the field declaration \texttt{code}, while \textit{Right} independently adds the field declaration \texttt{age}. 
Despite these changes not being conflicting, \jdime{} incorrectly reports an \textit{insert/insert} conflict. 
This occurs because, during matching, \jdime{} does not assign unique identifiers to field declarations and relies solely on its structural matching algorithm, which assigns a partial matching between the two properties--- as they have the same node kind. 
In contrast, \lastmerge{} properly treats the field name as a unique identifier and never matches nodes with different identifiers, thus correctly identifying the additions as non-conflicting, resulting on a clean merge.
\jdime's configuration could, and should, be adjusted to avoid this issue; it's not a fundamental limitation of the tool, or one that is inherent to language-specific tools.

Similar configuration differences also cause \lastmerge{} to report false positives due to method renaming. 
For example, in Figure~\ref{fig:afp_jdime_last_merge}, \textit{Left} modifies only the implementation of the method \texttt{greet}, while \textit{Right} changes both its body and signature by adding an argument. 
Since \lastmerge{} matches method declarations only when their signatures (name and argument types) are identical, it fails to match the different versions of \texttt{greet} from \textit{Left} and \textit{Right}. 
Instead, it mistakenly interprets \textit{Right}'s change as the addition of a new method \texttt{greet(String greet)} and the removal of the original \texttt{greet()}. 
Furthermore, because \lastmerge{} detects that \textit{Left} also modifies the original method, it classifies this as a \textit{modify/delete} conflict. 
In contrast, \jdime{} matches methods by name and subtree structure, correctly identifying the correspondence between the renamed methods and producing a clean merge.
Renaming conflicts are a common source of \afps{} in structured merge tools \cite{cavalcanti2017,lebenich2017}. Approaches to address this issue typically involve modifying general aspects of the matching process, which could be replicated in \lastmerge{}, without prejudicing language independence.

\begin{figure*}
    \centering
    \includegraphics[width=0.9\textwidth]{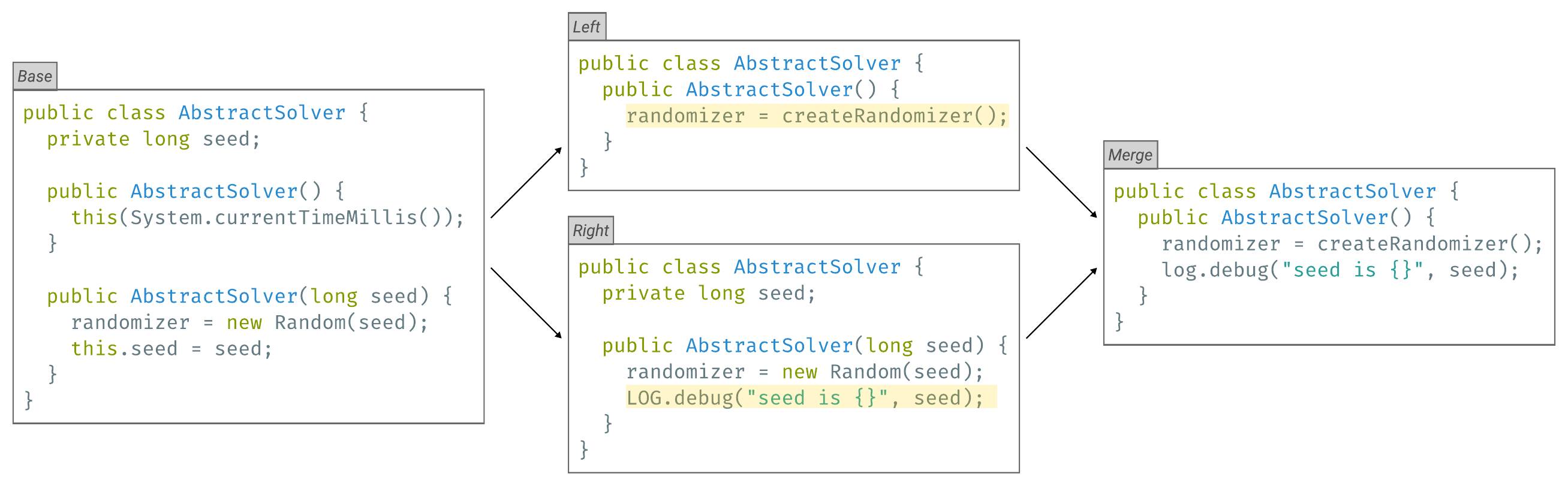}
    \caption{Merge scenario illustrating changes that lead into a false negative in \jdime{}. Changes are highlighted in yellow.}
    \label{fig:jdime_afn_overloads}
\end{figure*}

Turning to false negatives, the conflicts missed exclusively by \jdime{} (\afns{}) also arise from differences in matching configuration. 
To illustrate, consider the example in Figure~\ref{fig:jdime_afn_overloads}. 
The scenario involves independent changes to different constructors of the class \textit{AbstractSolver}. 
\textit{Left} retains only the no-argument constructor \texttt{AbstractSolver()}, modifies its body, and removes the field declaration \texttt{seed}. 
Meanwhile, \textit{Right} keeps the constructor \texttt{AbstractSolver(long seed)} and modifies its body by adding a new logging statement. 
These are conflicting changes, as each parent modifies constructors that were removed in the other revision. 
Since \jdime{} matches constructors based only on their name--- not their full signature--- it incorrectly matches the different constructors of \textit{Left} and \textit{Right}, merging both without reporting conflicts. 
However, due to the removal of the field declaration \texttt{seed}, the generated file cannot be compiled due to a missing symbol error. 
In contrast, \lastmerge{} treats each constructor as distinct nodes, correctly detecting and reporting the conflict.
jDime could be configured to match constructors by their full signature instead, which would result in the same behavior observed on \lastmerge{}; once again this highlights that the differences in output observed arise  from different configurations rather than language specific concerns.

We observe that \jdime{} misses actual conflicts when merging generic type arguments, resulting in compilation errors. 
Similarly, \lastmerge{} produces compilation errors when integrating \texttt{throws} declarations in method signatures. 
Both issues arise from incorrect configuration of children node ordering: \jdime{} treats generic type arguments as unordered, while \lastmerge{} treats \texttt{throws} declarations as ordered.
Finally, such differences in configuration can be adjusted, and do not reflect fundamental limitations of the tools.
Indeed, such configuration differences are not related to the language independence of \lastmerge{} in any sense.

\begin{mdframed}[style=mpdframe]
    \jdime{} and \lastmerge{} disagree on the existence of conflicts in 7.53\% of the scenarios. \lastmerge{} has fewer \afps{} than \jdime{}, but exhibits nearly three times more \afns{}. Our manual analysis shows that these discrepancies occur mainly due to implementation details and configuration differences, rather than to the language independent aspects of \lastmerge{}.
\end{mdframed}

\subsection*{\mergiraf{} and \spork{}}

Table~\ref{tab:spork_mergiraf_afps_afns} shows that \mergiraf{} reports almost twice as many \afps{} as \spork{}. 
Conversely, \mergiraf{} produces significantly fewer \afns{}.
Contrasting with the previous section, here the generic tool has more \afps{} but fewer \afns{}.
Similar to the previous section, we observe that the differences are mostly not due to the language independent aspects of the generic tool.

\begin{table}[htpb]
    \caption{Comparison of added false positives (\afps{}) and added false negatives (\afns{}) between \mergiraf{} and \spork{}.}
    \centering
    \begin{tabular}{ccc}
        \hline
        \textbf{Situation} & \textbf{\spork{}} & \textbf{\mergiraf} \\
        \hline
        Added false positives (\afps{}) & 150 & 290 \\
        \hline
        Added false negatives (\afns{}) & 107 & 62 \\
        \hline
    \end{tabular}
    \label{tab:spork_mergiraf_afps_afns}
\end{table}

Our manual analysis indicates the design decision of \mergiraf{} to use auto-tuning strongly influences the differences observed between the tools. 
With this strategy, \mergiraf{} first attempts an unstructured merge of the revisions and falls back to a structured approach only when conflicts arise. 
In contrast, \spork{} always applies a structured merge algorithm. 
We observe that, in 4 out of 5 scenarios analyzed, \spork{} failed to reproduce clean merges that were previously achieved by \mergiraf{} through unstructured merge. 
These failures typically result from incorrect or missing node matchings, which lead \spork{} to report spurious conflicts (\afps{}) in cases where \mergiraf{} produces conflict free results.
Adding auto-tunnig to \spork{} would be trivial, though; conversely, modifying \mergiraf{} to skip auto-tunning can also be easily achieved.

\begin{figure}
    \centering
    \includegraphics[width=0.5\textwidth]{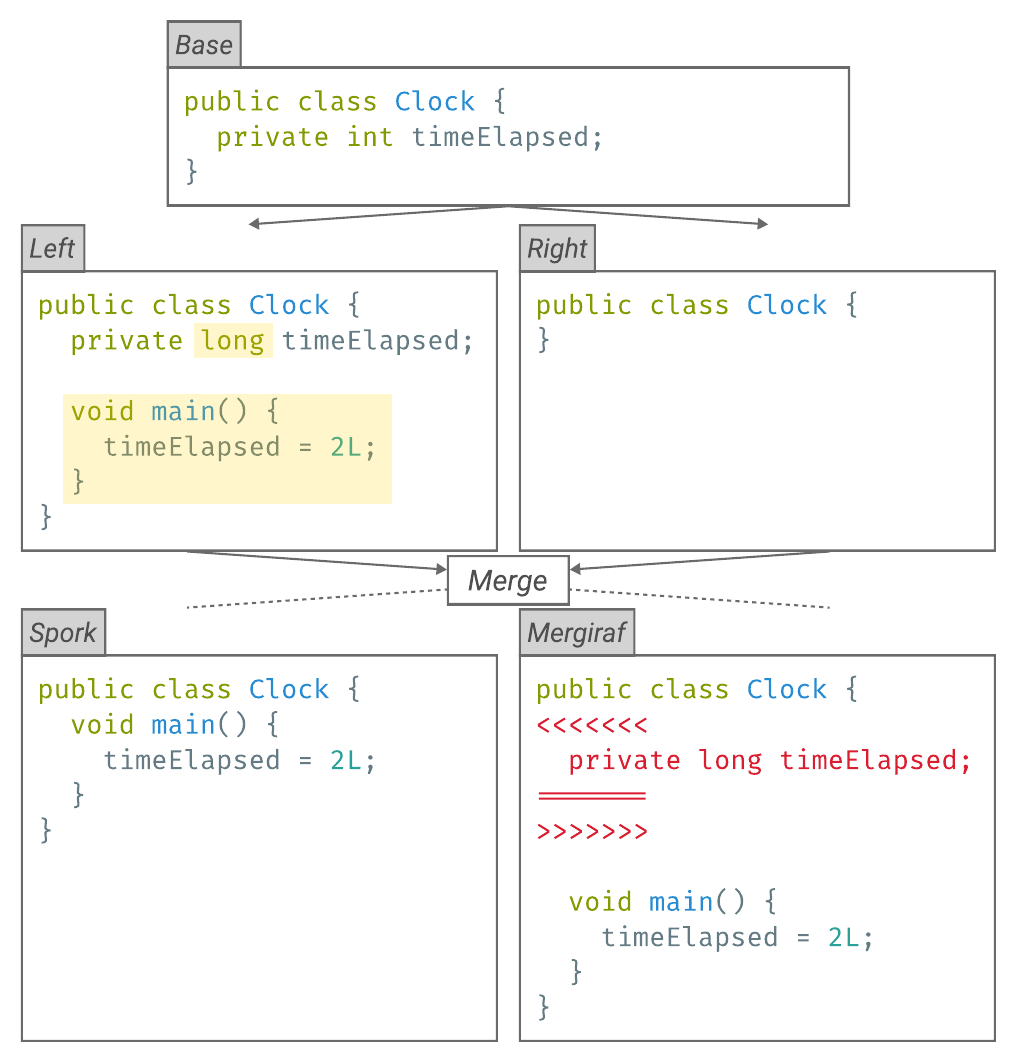}
    \caption{A merge scenario illustrating a \textit{delete/edit} conflict not detected by \spork{}. Changes are highlighted in yellow.}
    \label{fig:mergiraf_delete_edit}
\end{figure}

Algorithmic differences also contribute to the discrepancies observed between \mergiraf{} and \spork{}.
To illustrate this, consider the situation in Figure~\ref{fig:mergiraf_delete_edit}. 
\textit{Left} modifies the type of the field declaration \texttt{timeElapsed}, while \textit{Right} removes its declaration.
This situation, where one revision deletes a node modified by the other, characterizes a \textit{delete/edit} conflict.
\mergiraf{} and \spork{} use the same merge algorithm, whose original implementation is not able to detect such conflicts \cite{larsen2023}.
Instead, it silently deletes the node without reporting a conflict.
This way, \spork{} considers only the removal of the field declaration \texttt{timeElapsed} by right, and completely ignores the changes made by \textit{Left} to the same node. 
However, as \textit{Left} also adds a new reference to \texttt{timeElapsed} in the \texttt{main} method, the resulting file fails to compile due to a missing symbol error.
In contrast, \mergiraf{} extends the algorithm to keep track of deletions during the reconstruction of the merged tree.
After the merged tree is fully constructed, it checks wether one of the deleted nodes was modified on the other revision, enabling it to correctly detect and report the conflict.
Such extension is not a fundamental change in the algorithm, but rather an improvement to the original algorithm, which could be applied to \spork{} as well.

One can notice, however, that if \textit{Left} did not introduce a new reference to the \texttt{timeElapsed} property, the file produced by \spork{} would remain semantically valid and compile without errors. 
In general, \textit{delete/edit} conflicts only lead to issues when the changes introduced interfere semantically, thus leading into build-time errors \cite{dasilva2022}. 
Therefore, while \mergiraf{} adopts a more conservative strategy by always reporting \textit{delete/edit} conflicts, this choice may result in a higher number of \afps{}, as the scenarios where such conflicts truly affect the correctness of the merged program are relatively specific.

\begin{mdframed}[style=mpdframe]
    \spork{} and \mergiraf{} disagree on the existence of conflicts in 12.22\% of the scenarios. \mergiraf{} has nearly twice as many \afps{} than \spork{}, but exhibits significantly fewer \afns{}. Our manual analysis shows that these discrepancies occur mainly due to differences on the usage of auto-tuning and \mergiraf{} improvements to the original \spork{} algorithm, rather than the language independent aspects of the generic tool.
\end{mdframed}

\subsection{How \emph{generic} structured merge impacts merge runtime performance? (RQ2)}

\begin{figure}[H]
    \centering
    \includegraphics[width=0.45\textwidth]{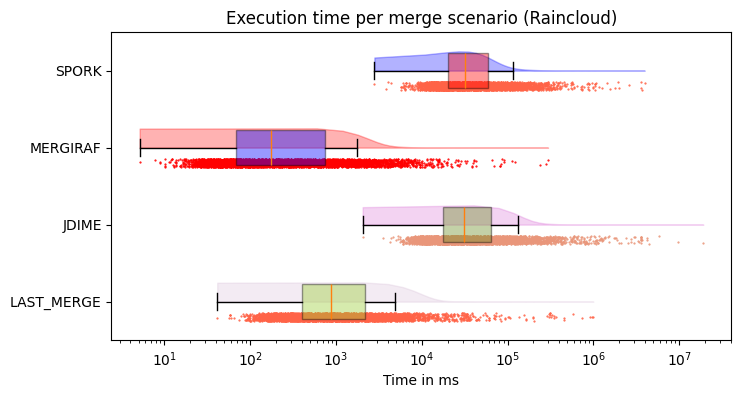}
    \caption{A raincloud plot displaying runtime execution per merge scenario for each tool. Each dot represents the average of 9 sequential executions of each tool in the same merge scenario. Time is in logarithmic scale.}
    \label{fig:merge_time}
\end{figure}

Figure~\ref{fig:merge_time} presents the runtime performance of the tools analyzed by merging the scenarios in our dataset. 
Overall, both \lastmerge{} and \mergiraf{} outperform \jdime{} and \spork{}, achieving speedups of at least one order of magnitude on average. 
These differences, however, arise mainly from implementation and design choices rather than fundamental algorithmic improvements or the language independent aspects of the generic tools.

Both \lastmerge{} and \mergiraf{} are implemented in Rust, whereas \spork{} and \jdime{} use Java. Rust is a systems programming language that compiles directly to machine code, enabling efficient execution. 
In contrast, Java compiles to to an intermediate representation (bytecode) that is later interpreted and executed on the Java Virtual Machine. 
Furthermore, while Java achieves memory safety in runtime by using garbage collection, Rust takes a unique ownership model that enforces memory safety in compile time, eliminating runtime overhead. 
Since structured merge tools perform CPU-intensive algorithms and numerous in-memory, large tree manipulation operations, their runtime performance is highly sensitive to the characteristics of the implementation language.

Beyond the choice of programming language, other factors also influence performance differences. For example, \mergiraf{} employs techniques known to enhance efficiency, such as auto-tuning~\cite{apel2012}. 
In this approach, the tool first attempts an unstructured merge and resorts to structured methods only if conflicts arise. 
Invoking structured merge selectively leads to significant performance gains~\cite{seibt2021}. 
Notably, an auto-tuning strategy could also be implemented without further effort, potentially yielding similar benefits for the other tools analyzed.

Finally, even if we focus our comparison on \lastmerge{}--- which does not use auto-tuning--- and assume a conservative 10x performance penalty when comparing Java to Rust implementations, based on prior work~\cite{couto2017}, \lastmerge{} still delivers performance comparable to state of the art tools, integrating 80.2\% of the scenarios in less than three seconds. 
This suggests that generic structured merge does not impose a computational cost that is prohibitive compared to existing tools, and can be used in most situations in industry.

\subsection{Threats to validity}
\label{sec:threats_to_validity}

Our \afps{} and \afns{} analysis (see Section~\ref{sec:methodology}) relies on a heuristic that combines static and semantic analysis to approximate the existence of conflicts. 
Using the merge commit in the repository is a good approximation of the expected merge result, but developers might, for instance, have accepted the commit and immediately after noted a problem, later fixed it in a subsequent commit.
Failing builds are a quite robust approximation of problem in the result yielded by the merge tool, but test is as robust as the project test suite itself.
Nevertheless, this criteria has been used in recent work~\cite{schesch2024} and is stronger than the ones used in previous work~\cite{fengmin2018,fengmin2019,cavalcanti2017}, which focus only on the first part of our criteria.  

Relying on manual analysis for identifying the reasons for false positives and false negatives can be challenging. 
Accurately classifying such scenarios require a deep understanding of the project implementation and even the original developers may sometimes overlook conflicting changes. 
Instead, our manual analysis focuses on understanding the underlying reasons for the differences in results produced by the tools. 
More specially, we focus on understanding wether these differences arise because of the introduction of language independence aspects in the generic tools. 
This approach is less error-prone, as the authors possess in-depth knowledge of the design and behavior of each tool. 
Additionally, all findings were thoroughly discussed among the authors to ensure accurate interpretations.

Our sample represents only a subset of possible merge scenarios. 
Specifically, we focus on cases extracted from publicly available Java projects on GitHub, which may not capture the full spectrum of programming practices. 
To mitigate this limitation, we rely on a comprehensive dataset introduced by the work of Schesch et al.~\cite{schesch2024}, which includes a diverse collection of projects spanning various domains and development practices, and can be considere the state of the art dataset for merge tool studies.

Finally, despite the language independence of \lastmerge{} and \mergiraf{}, our findings may still be specific to Java, the programming language used in our evaluation. 
Since \jdime{} and \spork{} support only Java, generalizing our results to other languages would need structured merge tools for other languages, but these are hardly available. 
Additional studies comparing \lastmerge{} and \mergiraf{} with language specific structured merge tools across different languages are necessary to better assess their applicability in broader contexts.

\section{Related Work}
\label{sec:related_work}

Software merging is a well studied problem, and many approaches have been proposed to address it.

\subsection{Structured Merge}

Several structured merge tools have been proposed previously. These tools rely on the syntax and semantics of specific programming languages to perform the merge, in contrast to \lastmerge{}, which is designed to be more flexible and not limited to any particular language.

Apel et al. \cite{apel2012} introduced \jdime{}, a tool that applies a structured merge strategy to integrate Java programs. \jdime{} employs auto-tuning to dynamically switch between structured and unstructured merge based on the presence of conflicts. Later, Leßenich et al. \cite{lebenich2017} enhanced \jdime{} by incorporating a look-ahead mechanism into the node matching algorithm. This improvement enables the tool to match nodes across different hierarchical levels, allowing it to better detect refactorings such as renamings and code movements. As a result, matching precision increased by 28\% without compromising performance.

Zhu et al. \cite{fengmin2019} proposed \textsc{AutoMerge}. Built on top of \jdime{}, it introduces a new matching heuristic that aims to quantify the quality of a matching between the nodes. This heuristic quantifies the similarity between nodes, and the algorithm seeks to maximize this quality function to avoid incorrect or unrelated matchings, thereby improving precision. Their evaluation shows that \textsc{AutoMerge} more closely reproduces the developer resolved merges found in commit histories compared to \jdime{}, at the expense of a performance overhead.

Larsen et al. \cite{larsen2023} proposed \spork{}, a structured merge tool for Java. \spork{} relies on GumTree \cite{gumtree} to compute node matchings and introduces a \textit{high-fidelity pretty-printing} mechanism that reuses code fragments from the original revisions. This enables the preservation of source code formatting, and their evaluation shows that \spork{} retained the original source code in over 90\% of merged files when compared to \jdime{}.

Despite their benefits, extending existing tools to support other programming languages is challenging, primarily due to their strong coupling with the specific languages for which they were originally designed. In contrast, \lastmerge{} operates on generic tree structures and leverages the \treesitter{} infrastructure to parse source code into these trees. Additionally, \lastmerge{} provides an interface that allows users to extend the tool for use with other programming languages.

\subsection{Other approaches}

Other techniques have been explored as alternatives to structured merge. One such approach is semistructured merge, a hybrid strategy designed to balance precision and accuracy. Semistructured merge applies structured merge to higher-level elements, such as method declarations, while utilizing unstructured merge for lower-level elements, like expressions and statements within method bodies.

Apel et al. \cite{apel2011} proposed \textsc{FSTMerge}, a generic semistructured merge engine. They have configured it for use with C\#, Java, and Python. To support a new language, developers must provide an annotated grammar that specifies the elements of the language where the order does not matter. However, this approach requires a deep understanding of the language's grammar and the ability to annotate it correctly, which can pose a barrier for many users. In contrast, \lastmerge{} relies on the \treesitter{} parser framework, which has been instantiated for over 350 languages, and offers a thin configuration interface for extending the tool for usage with new languages. This makes \lastmerge{} more flexible and easier to adapt to new languages compared to \textsc{FSTMerge}.

Cavalcanti et al. \cite{cavalcanti2017} evaluated \textsc{FSTMerge} by comparing it with unstructured merge on Java code. They found that \textsc{FSTMerge} significantly reduces the number of false positives, but this comes at the expense of missing actual conflicts (false negatives). Based on their findings, they proposed \textsc{s3m}, a semistructured merge tool for Java built on top of \textsc{FSTMerge}. \textsc{s3m} extends \textsc{FSTMerge} by introducing \textit{handlers} that update the final tree using heuristics based on information gathered during the merge process. These \textit{handlers} enhance merge accuracy by addressing cases where \textsc{FSTMerge} previously produced additional false positives and false negatives, such as in renaming scenarios. We draw inspiration from \textsc{s3m} to implement the \textit{parsing handlers} in \lastmerge{}.

Finally, Cavalcanti et al. \cite{cavalcanti2024} proposed \textsc{Sesame}, a tool designed to emulate structured merge by leveraging semistructured merge along with language-specific syntactic separators. The tool uses semistructured merge to integrate revisions; however, before performing unstructured merge on lower-level elements, it synthetically splits the source code based on these language separators, such as curly braces in Java. This splitting infers a structure that, when combined with unstructured merge, emulates structured merge. Their empirical evaluation demonstrates that \textsc{Sesame} achieves results comparable to those of structured tools, positively by reducing false positives and negatively by increasing the number of false negatives.

Despite being a structured merge tool, \lastmerge{} can emulate the semistructured behavior of both \textsc{FSTMerge}, \textsc{s3m} and \textsc{Sesame}.
This is possible because the user can configure the tool to stop parsing at an intermediate tree level, treating all descendants of that node as a single textual element.
\section{Conclusion}
\label{sec:conclusion}

In this paper, we introduced \lastmerge{}, a \emph{generic} structured merge tool.
Our goal was to reduce the barriers that hinder the adoption of structured merge in industry, where support for multiple languages is often needed for most nontrivial projects.
\lastmerge{} relies on a core merge engine that operates over generic trees.
Combined with a high level description of language specific aspects, developers can easily adapt \lastmerge{} for new languages, or refine support for existing ones.

Our comparative analysis between \emph{generic} and language specific tools show no evidence of significant impacts on both merge accuracy and runtime performance.
These results suggest that \emph{generic} structured merge tools can effectively replace language-specific ones, achieving similar levels of accuracy and efficiency, thus paving the way for broader adoption of structured merge in industry.
Future work will focus on extending language support and enhancing \lastmerge{} to further improve merge accuracy.

\bibliographystyle{ieeetr}
\bibliography{references}

\end{document}